\documentclass[authoryear]{elsarticle}

\begin{document}

\newcommand{\LCDM}{$\Lambda$CDM}
\newcommand{\azero}{\ensuremath{\mathrm{a}_{0}}}
\newcommand{\ML}{\ensuremath{M_*/L}}
\newcommand{\kms}{\ensuremath{\mathrm{km}\,\mathrm{s}^{-1}}}
\newcommand{\galunits}{\ensuremath{\mathrm{km}\,\mathrm{s}^{-1}\,\mathrm{kpc}^{-1}}}
\newcommand{\galacc}{\ensuremath{\mathrm{km}^2\,\mathrm{s}^{-2}\,\mathrm{kpc}^{-1}}}
\newcommand{\MLsun}{\ensuremath{\mathrm{M}_{\odot}/\mathrm{L}_{\odot}}}
\newcommand{\Lsun}{\ensuremath{\mathrm{L}_{\odot}}}
\newcommand{\Msun}{\ensuremath{\mathrm{M}_{\odot}}}

\title{Testing Galaxy Formation and Dark Matter \\ with Low Surface Brightness Galaxies}

\author[]{Stacy S. McGaugh}
\address[]{Department of Astronomy, Case Western Reserve University, \\ 10900 Euclid Avenue, Cleveland, OH 44106, USA}

\begin{abstract}
Galaxies are the basic structural element of the universe; galaxy formation theory seeks to explain how these structures came to be. 
I trace some of the foundational ideas in galaxy formation, with emphasis on the need for non-baryonic cold dark matter.
Many elements of early theory did not survive contact with observations of low surface brightness galaxies,
leading to the need for auxiliary hypotheses like feedback.
The failure points often trace to the surprising predictive successes of an alternative to dark matter, the Modified Newtonian Dynamics (MOND).
While dark matter models are flexible in accommodating observations, they do not provide the predictive capacity of MOND.
If the universe is made of cold dark matter, why does MOND get any predictions right?
\end{abstract}


\maketitle

\section{Cosmic Context}
\label{sec:intro}

Cosmology is the science of the origin and evolution of the universe: the biggest of big pictures.
The modern picture of the hot big bang is underpinned by three empirical pillars: an expanding universe (Hubble expansion),
Big Bang Nucleosynthesis (BBN: the formation of the light elements through nuclear reactions in the early universe), and the relic
radiation field (the Cosmic Microwave Background: CMB) \citep{Peebles,Harrison}. The discussion here will take this framework
for granted. 

The three empirical pillars fit beautifully with General Relativity (GR).
Making the simplifying assumptions of homogeneity and isotropy, Einstein's equations can be applied to treat the entire universe
as a dynamical entity. As such, it is compelled either to expand or contract. Running the observed expansion
backwards in time, one necessarily comes to a hot, dense, early phase. 
This naturally explains the CMB, which marks the transition from an opaque plasma to a transparent gas \citep{WeissARAA,SZARAA}.
The abundances of the light elements can be explained in detail with BBN provided the universe expands in the first few minutes as predicted
by GR when radiation dominates the mass-energy budget of the universe \citep{BBNARAA}. 

The marvelous consistency of these early universe results with the expectations of GR builds confidence that the hot big bang is the
correct general picture for cosmology. It also builds overconfidence that GR is completely sufficient to describe the universe. 
Maintaining consistency with modern cosmological data is only possible with the addition of two auxiliary hypotheses:
dark matter and dark energy. These invisible entities are an absolute requirement of the current version of the most-favored
cosmological model, \LCDM. The very name of this model is born of these dark materials: $\Lambda$ is Einstein's cosmological constant,
of which `dark energy' is a generalization, and CDM is cold dark matter. 

Dark energy does not enter much into the subject of galaxy formation. It mainly helps to set the background cosmology in which galaxies
form, and plays some role in the timing of structure formation. This discussion will not delve into such details, and I note only that it
was surprising and profoundly disturbing that we had to 
reintroduce \citep[e.g.,][]{Lambda1990,YoshiiPeterson,OS,Riess1998,Perlmutter1999} Einstein's so-called `greatest blunder.'

Dark matter, on the other hand, plays an intimate and essential role in galaxy formation.
The term `dark matter' is dangerously crude, as it can reasonably be used to mean anything that is not seen.
In the cosmic context, there are at least two forms of unseen mass: normal matter that happens not to glow in a way that is easily seen ---
not all ordinary material need be associated with visible stars --- and non-baryonic cold dark matter. It is the latter
form of unseen mass that is thought to dominate the mass budget of the universe and play a critical role in galaxy formation. 

I discuss the basic historical motivations for non-baryonic CDM in \S \ref{sec:CDM}, and provide an overview of essential galaxy properties in \S \ref{sec:props}.
These are intended as a minimal introduction for the non-expert. In \S \ref{sec:galform}, I provide a broad overview of galaxy formation together with specific
criticisms of galaxy formation theory. In \S \ref{sec:MOND} I discuss some of the unexpected predictive successes of the alternative to dark matter, 
MOND \citep{milgrom83a}. The existential problem that this poses for the dark matter paradigm is addressed in \S \ref{sec:whylordwhy}. 

\section{Cold Dark Matter}
\label{sec:CDM}

Cold dark matter is some form of slow moving, non-relativistic (`cold') particulate mass
that is not composed of normal matter (baryons). Baryons are the family of particles that include protons and neutrons. As such, they compose the
bulk of the mass of normal matter, and it has become conventional to use this term to distinguish between normal, baryonic matter and
the non-baryonic dark matter. 

The distinction between baryonic and non-baryonic dark matter is no small thing.
Non-baryonic dark matter must be a new particle that resides in a new `dark sector' that is completely distinct from
the usual stable of elementary particles. We do not just need some new particle, we need one (or many) that reside
in some sector beyond the framework of the stubbornly successful Standard Model of particle physics. 
Whatever the solution to the mass discrepancy problem turns out to be, it requires new physics. 

The cosmic dark matter must be non-baryonic for two basic reasons. 
First, the mass density of the universe measured gravitationally \citep[$\Omega_m \approx 0.3$, e.g.,][]{FG79,Davis1980,EndCDM}
clearly exceeds the mass density in baryons as constrained by BBN \citep[$\Omega_b \approx 0.05$, e.g.,][]{BBN}. 
There is something gravitating that is not ordinary matter: $\Omega_m > \Omega_b$.

The second reason follows from the absence of large fluctuations in the CMB \citep{Silkdamping,PeeblesYu,SZARAA}. 
The CMB is extraordinarily uniform in temperature across the sky, varying by only $\sim 1$ part in $10^5$ \citep{COBE}. 
These small temperature variations correspond to variations in density. 
Gravity is an attractive force; it will make the rich grow richer. Small density excesses will tend to attract more mass,
making them larger, attracting more mass, and leading to the formation of large scale structures, including galaxies.
But gravity is also a weak force: this process takes a long time. 
In the long but finite age of the universe, gravity plus known baryonic matter does not suffice to go from the initially smooth, highly uniform
state of the early universe to the highly clumpy, structured state of the local universe \citep{Peebles}. The solution is to boost the process
with an additional component of mass --- the cold dark matter --- that gravitates without interacting with the photons, thus 
getting a head start on the growth of structure while not aggravating the amplitude of temperature fluctuations in the CMB.

Taken separately, one might argue away the need for dark matter. 
Taken together, these two distinct arguments convinced nearly everyone, including myself, of the absolute need for non-baryonic dark matter.
Consequently, CDM became established as the leading paradigm during the 1980s \citep{Peebles1984,origWIMP}.
The paradigm has snowballed since that time, the common attitude among cosmologists being that CDM \textit{has} to exist.

From an astronomical perspective, the CDM could be any slow-moving, massive object that does not interact with photons nor participate in BBN. 
The range of possibilities is at once limitless yet highly constrained.
Neutrons would suffice if they were stable in vacuum, but they are not. 
Primordial black holes are a logical possibility, but if made of normal matter, they must somehow form in the first second after the 
Big Bang to not impair BBN. At this juncture, microlensing experiments have excluded most plausible mass ranges that
primordial black holes could occupy \citep{BHmicrolensing}. It is easy to invent hypothetical dark matter candidates, but difficult 
for them to remain viable.

From a particle physics perspective, the favored candidate is a Weakly Interacting Massive 
Particle \citep[WIMP:][]{Peebles1984,origWIMP}. WIMPs are expected to be the lightest stable supersymmetric partner particle
that resides in the hypothetical supersymmetric sector \citep{SUSYprimer}. 
The WIMP has been the odds-on favorite for so long that it is often used synonymously with the more generic term `dark matter.'
It is the hypothesized particle that launched a thousand experiments. Experimental searches for WIMPs have matured over the
past several decades, making extraordinary progress in not detecting dark matter \citep{Xenon1T}.
Virtually all of the parameter space in which WIMPs had been predicted to reside \citep{Trotta2008} is now excluded.
Worse, the existence of the supersymmetric sector itself, once seemingly a sure thing, remains entirely hypothetical, 
and appears at this juncture to be a beautiful idea that nature declined to implement. 

In sum, we must have cold dark matter for both galaxies and cosmology, but we have as yet no clue to what it is.

\section{Galaxy Properties}
\label{sec:props} 

Cosmology entered the modern era when \citet{Hubble1929} resolved the debate over the nature of spiral nebulae by
measuring the distance to Andromeda, establishing that vast stellar systems --- galaxies --- exist external to and coequal with the Milky Way. 
Galaxies are the primary type of object observed when we look beyond the confines of our own Milky Way:
they are the building blocks of the universe.
Consequently, galaxies and cosmology are intertwined: it is impossible to understand one without the other.

\begin{figure}[t]
\begin{center}
\includegraphics[width=1.0\textwidth]{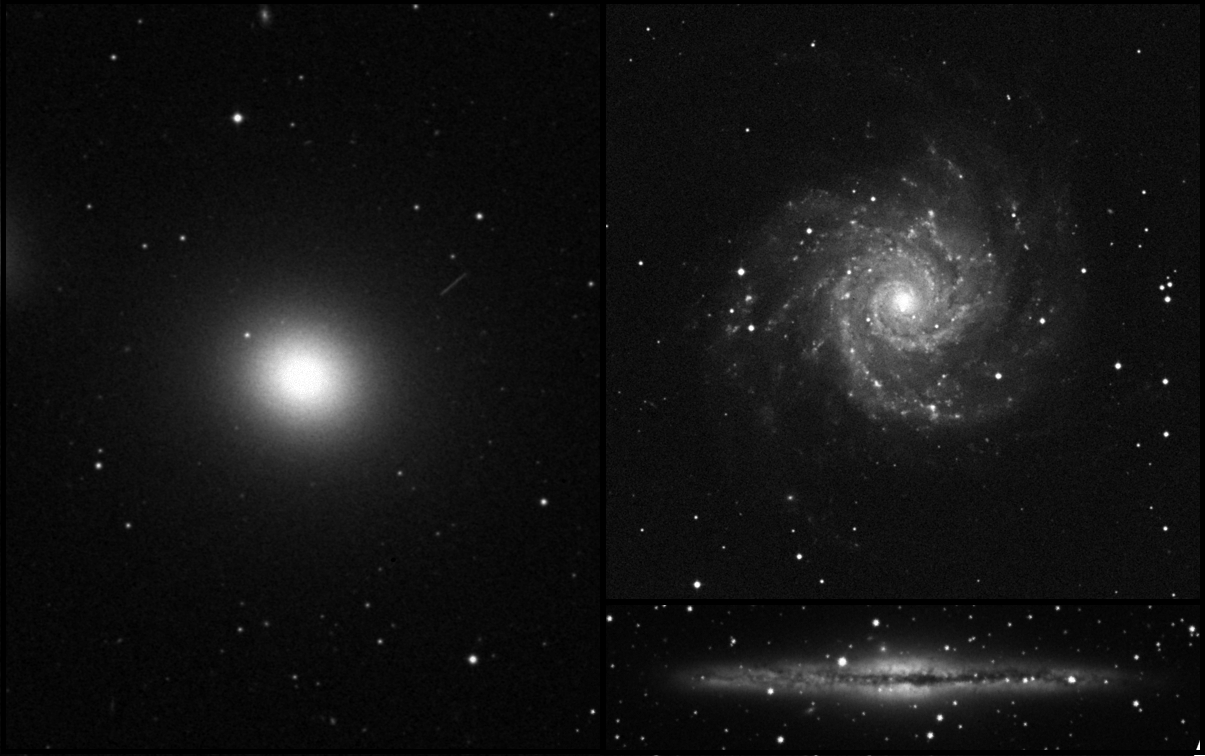}
\end{center}
\caption{Galaxy morphology. These examples shown an early type elliptical galaxy (NGC 3379, left), 
and two late type disk galaxies: a face-on spiral (NGC 628, top right), and an edge-on disk galaxy (NGC 891, bottom right).
Elliptical galaxies are quasi-spherical, pressure supported stellar systems that
tend to have predominantly old stellar populations, usually lacking young stars or 
much in the way of the cold interstellar gas from which they might form.
In contrast, late type galaxies (spirals and irregulars) are thin, rotationally supported disks.
They typically contain a mix of stellar ages and cold interstellar gas from which new stars continue to form.
Interstellar dust is also present, being most obvious in the edge-on case. 
Images from Palomar Observatory, Caltech.
}
\label{fig:NGC3379N628NGC821}
\end{figure}

Here I sketch a few essential facts about the properties of galaxies.
This is far from a comprehensive list \citep[see, for example][]{BT} 
and serves only to provide a minimum framework for the subsequent discussion.
The properties of galaxies are often cast in terms of morphological type, starting with Hubble's tuning fork diagram.
The primary distinction is between Early Type Galaxies (ETGs) and Late Type Galaxies (LTGs), which is a matter of basic structure.
ETGs, also known as elliptical galaxies, are three dimensional, ellipsoidal systems that are pressure supported: 
there is more kinetic energy in random motions than in circular motions, a condition described as dynamically hot.
The orbits of stars are generally eccentric and oriented randomly with respect to one another, filling out the ellipsoidal shape
seen in projection on the sky. LTGs, including spiral and irregular galaxies, are thin, quasi-two dimensional,
rotationally supported disks. The majority of their stars orbit in the same plane in the same direction on low eccentricity orbits.
The lion's share of kinetic energy is invested in circular motion, with only small random motions, a condition described as
dynamically cold. Examples of early and late type galaxies are shown in Fig.\ \ref{fig:NGC3379N628NGC821}.

Finer distinctions in morphology can be made within the broad classes of early and late type galaxies, 
but the basic structural and kinematic differences suffice here. The disordered motion of ETGs is a natural consequence of 
violent relaxation \citep{violentrelax} in which a stellar system reaches a state of dynamical equilibrium from a
chaotic initial state. This can proceed relatively quickly from a number of conceivable initial conditions, and is a rather natural
consequence of the hierarchical merging of sub-clumps expected from the Gaussian initial conditions indicated by observations
of the CMB \citep{WhiteVR}. 
In contrast, the orderly rotation of dynamically cold LTGs requires a gentle settling of gas into a rotationally supported disk.
It is essential that disk formation occur in the gaseous phase, as gas can dissipate and settle to the preferred plane specified by
the net angular momentum of the system. 
Once stars form, their orbits retain a memory of their initial state for a period typically much greater than the age of the universe \citep{BT}. 
Consequently, the bulk of the stars in the spiral disk must have formed there after the gas settled.

In addition to the dichotomy in structure, ETGs and LTGs also differ in their evolutionary history.
ETGs tend to be `red and dead,' which is to say, dominated by old stars. 
They typically lack much in the way of recent star formation,
and are often devoid of the cold interstellar gas from which new stars can form. 
Most of their star formation happened in the early universe, and may have involved the merger of multiple protogalactic fragments.
Irrespective of these details, massive ETGs appeared early in the universe \citep{impossiblyearly}, 
and for the most part seem to have evolved passively since \citep{Franckmass}. 

Again in contrast, LTGs have on-going star formation in interstellar media replete with cold atomic and molecular gas. 
They exhibit a wide range in stellar ages, from newly formed stars to ancient stars dating to near the beginning of time.
Old stars seem to be omnipresent, famously occupying globular clusters but also present in the general disk population.
This implies that the gaseous disk settled fairly early, though accretion may continue over a
long timescale \citep{Gdwarfprob,GdwarfHW}. 
Old stars persist in the same orbital plane as young stars \citep{BM}, which precludes much subsequent merger activity,  
as the chaos of merging distorts orbits. Disks can be over-heated \citep{TothO} and transformed
by interactions between galaxies \citep{TT72}, even turning into elliptical galaxies during major mergers \citep{BarnesHernquist}.

Aside from its morphology, an obvious property of a galaxy is its mass. Galaxies exist over a large range
of mass, with a type-dependent characteristic stellar mass of $5 \times 10^{10}\;\Msun$ for disk dominated systems
\citep[the Milky Way is very close to this mass:][]{BHGreview} and $10^{11}\;\Msun$ for elliptical galaxies \citep{GAMAmassfcn}. 
Above this characteristic mass, the number density of galaxies declines sharply, though individual galaxies exceeding 
a few $10^{11}\;\Msun$ certainly exist. The number density of galaxies increases gradually to lower masses, with no known minimum.
The gradual increase in numbers does not compensate for the decrease in mass: integrating over the distribution,
one finds that most of the stellar mass is in bright galaxies close to the characteristic mass.

\begin{figure}[t]
\begin{center}
\includegraphics[width=1.0\textwidth]{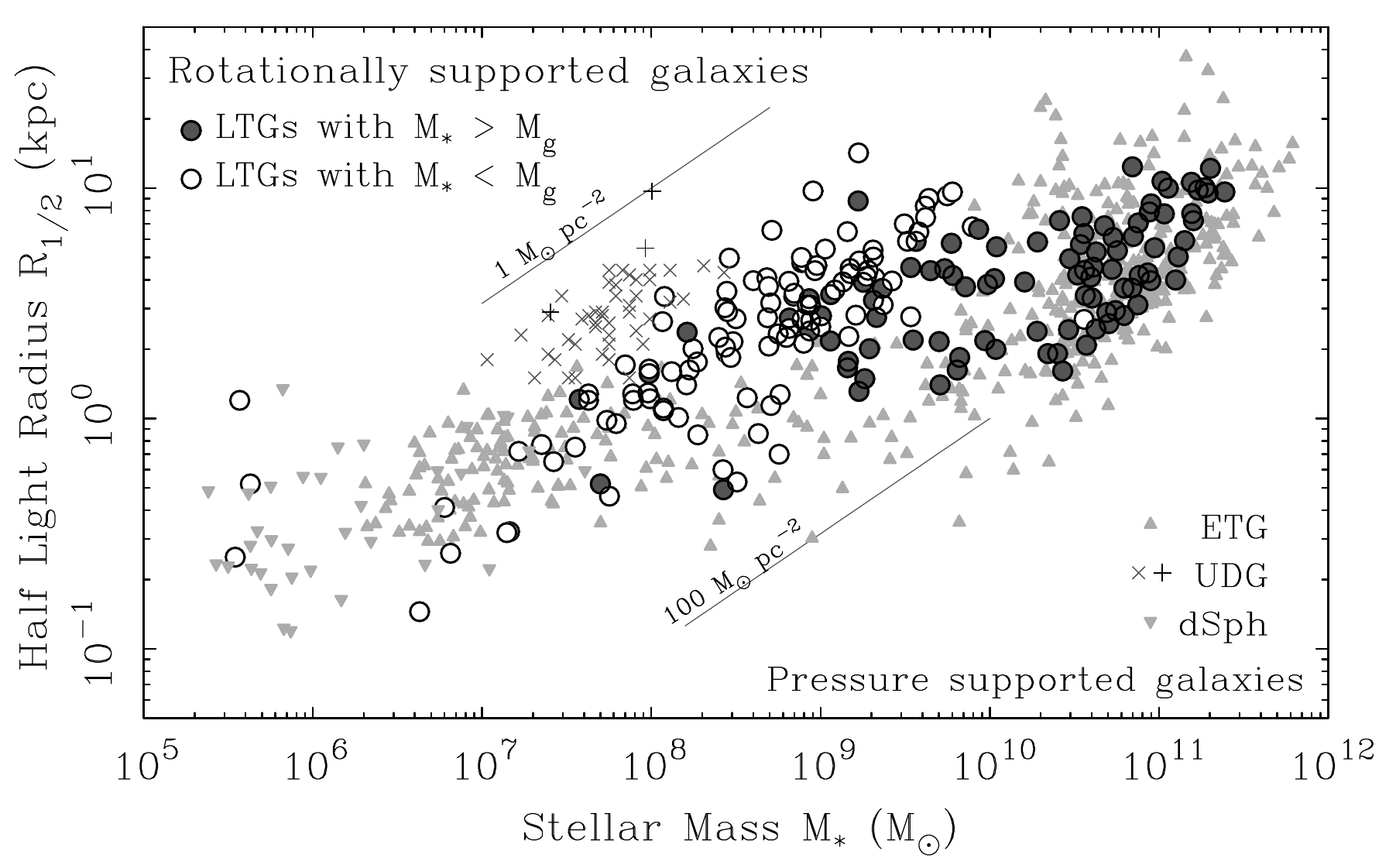}
\end{center}
\caption{Galaxy size and mass. The radius that contains half of the light is plotted against the stellar mass.
Galaxies exist over many decades in mass, and exhibit a considerable variation in size at a given mass.
Early and late type galaxies are demarcated with different symbols, as noted.
Lines illustrate tracks of constant stellar surface density. The data for 
ETGs are from the compilation of \citet{DFR} augmented by dwarf Spheroidal (dSph) galaxies in the Local Group compiled by
\citet{OneLaw}. Ultra-diffuse galaxies \citep[UDGs:][$\times$ and $+$, respectively]{vdKUDG,mihosUDG} 
have unsettled kinematic classifications at present, but most seem likely to be pressure supported ETGs.
The bulk of the data for LTGs is from the SPARC database \citep{SPARC}, augmented by cases that are noteworthy for their
extremity in mass or surface brightness \citep{MB94,dBvdH95,Dalc97,LSB2011,LeoPphot,ChrisCarr,ComaP}.
The gas content of these star-forming systems adds a third axis, illustrated crudely here by whether an LTG is made
more of stars or gas (filled and open symbols, respectively).
}
\label{fig:sizemass}
\end{figure}

Galaxies have a characteristic size and surface brightness. The same amount of stellar mass can be concentrated in a
high surface brightness (HSB) galaxies, or spread over a much larger area in a low surface brightness (LSB) galaxy.
For the purposes of this discussion, it suffices to assume that the observed luminosity is proportional to the mass
of stars that produces the light. Similarly, the surface brightness measures the surface density of stars. Of the
three observable quantities of luminosity, size, and surface brightness, only two are independent: the luminosity is
the product of the surface brightness and the area over which it extends. The area scales as the square of the linear size.

The distribution of size and mass of galaxies is shown in Fig.\ \ref{fig:sizemass}.
This figure spans the range from tiny dwarf irregular galaxies containing `only' a few hundred thousand stars
to giant spirals composed of hundreds of billions of stars with half-light radii ranging from hundreds of parsecs to tens of kpc.
The upper boundaries represent real, physical limits on the sizes and masses of galaxies.
Bright objects are easy to see; if still higher mass galaxies were common, they would be readily detected and cataloged. 
In contrast, the lower boundaries are set by the limits of observational sensitivity (``selection effects''): 
galaxies that are physically small or low in surface brightness are difficult to detect and
are systematically under-represented in galaxy catalogs \citep{Disney,AllenShu,MBS95}.

Individual galaxies can be early type or late type, high mass or low mass, large or small in linear extent, high or low surface
brightness, gas poor or gas rich. No one of these properties is completely predictive of the others: the correlations that do
exist tend to have lots of intrinsic scatter. The primary exception to this appears to involve the kinematics. 
Massive galaxies are fast rotators; low mass galaxies are slow rotators. This Tully-Fisher relation \citep{TForig} is 
one of the strongest correlations in extragalactic astronomy \citep{SPARCTF}. It is thus necessary to simultaneously explain both
the chaotic diversity of galaxy properties and the orderly nature of their kinematics \citep{IAU_review}.

Galaxies do not exist in isolation. Rather than being randomly distributed throughout the universe, they tend to cluster together:
the best place to find a galaxy is in the proximity of another galaxy \citep{VeraThesis}. 
A common way to quantify the clustering of galaxies is the two-point correlation function $\xi(r)$ \citep{LSSbook}.
This measures the excess probability of finding a galaxy within a distance $r$ of a reference galaxy relative to a random distribution. 
The observed correlation function is well approximated as a power law whose slope and normalization varies with galaxy population.
ETGs are more clustered than LTGs, having a longer correlation length: $r_0 \approx 9$ Mpc for red galaxies vs.\ $\sim 5$ Mpc for blue
galaxies \citep{Zehavi2011}. Here we will find this quantity to be of interest for comparing the distribution of high and low surface brightness galaxies.

\section{Galaxy Formation}
\label{sec:galform}

Galaxies are gravitationally bound condensations of stars and gas in a mostly empty, expanding universe. 
The tens of billions of solar masses of baryonic material that 
comprise the stars and gas of the Milky Way now reside mostly within a radius of 20 kpc. At the average density of the universe,
the equivalent mass fills a spherical volume with a comoving radius a bit in excess of 1 Mpc. This is a large factor by which 
a protogalaxy must collapse, starting from the very smooth ($\sim 1$ part in $10^5$) 
initial condition at $z = 1090$ observed in the CMB \citep{Planck2018}. Dark matter --- in particular, non-baryonic cold dark matter ---
plays an essential role in speeding this process along. 

The mass-energy of the early universe is initially dominated by the radiation field. The baryons are held in thrall to the photons
until the expansion of the universe turns the tables and matter becomes dominant. Exactly when this happens depends on the
mass density \citep{LSSbook}; for our purposes it suffices to realize that the baryonic components of galaxies can not begin
to form until well after the time of the CMB. However, since CDM does not interact with photons, it is not subject to this limitation.
The dark matter can begin to form structures --- dark matter halos --- that form the scaffolding of future structure. 
Essential to the \LCDM\ galaxy formation paradigm is that the dark matter halos form first, seeding the subsequent formation
of luminous galaxies by providing the potential wells into which baryons can condense once free from the radiation field.

The theoretical expectation for how dark matter halos form is well understood at this juncture.
Numerical simulations of cold dark matter --- mass that interacts only through gravity in an expanding universe --- 
show that quasi-spherical dark matter halos form with a characteristic `NFW' \citep[e.g.,][]{NFW} density profile.
These have a `cuspy' inner density profile in which the density of dark matter increases towards the center 
approximately\footnote{The exact details of this profile have been widely debated, both theoretically and observationally.
These details are not relevant here.} as a power law, $\rho(r \rightarrow 0) \sim r^{-1}$. 
At larger radii, the density profile falls of as $\rho(r \rightarrow \infty) \sim r^{-3}$. 
The centers of these halos are the density peaks around which galaxies can form.

The galaxies that we observe are composed of stars and gas: normal baryonic matter.
The theoretical expectation for how baryons behave during galaxy formation is not well understood \citep{Aquila}.
This results in a tremendous and long-standing disconnect between theory and observation. 
We can, however, stipulate a few requirements as to what needs to happen.
Dark matter halos must form first; the baryons fall into these halos afterwards. 
Dark matter halos are observed to extend well beyond the outer edges of visible galaxies, so
baryons must condense to the centers of dark matter halos.
This condensation may proceed through both the hierarchical merging of protogalactic fragments 
(a process that has a proclivity to form ETGs) and the more gentle accretion of gas into rotating disks (a requirement to form LTGs). 
In either case, some fraction of the baryons form the observed, luminous component of a galaxy at the center of a CDM halo.
This condensation of baryons necessarily affects the dark matter gravitationally, with the net effect of dragging some of it
towards the center \citep{Blumenthal,Dubinski94,Oleg04,SM05}, thus compressing the dark matter halo from its initial condition
as indicated by dark matter-only simulations like those of \citet{NFW}.
These processes must all occur, but do not by themselves suffice to explain real galaxies.

Galaxies formed in models that consider only the inevitable effects described above suffer many serious defects.
They tend to be too massive \citep{Benson03,Abadi},
too small \citep[the angular momentum catastrophe:][]{Katz1992,Steinmetz99,dOnghia06}, 
have systematically too large bulge-to-disk ratios \citep[the bulgeless galaxy problem:][]{dOnghia04,Kormendy10},
have dark matter halos with too much mass at small radii \citep[the cusp-core problem:][]{MooreCuspCore,KdN08,KdN09,cuspcore,KdN14}, 
and have the wrong over-all mass function \citep[the over-cooling problem, e.g.,][]{overcooling}, also known locally as the missing 
satellite problem \citep{Kmissingsatellites,Mmissingsatellites}. 
This long list of problems have kept the field of galaxy formation a lively one:
there is no risk of it becoming a victim its own success through the appearance of one clearly-correct standard model.

\subsection{Threads of Development}
\label{sec:threads}

Entering the 1980s, options for galaxy formation were frequently portrayed as a dichotomy between monolithic galaxy formation \citep{ELS} and
the merger of protogalactic fragments \citep{SZ}. The basic idea of monolithic galaxy formation is that the initial $\sim 1$ Mpc cloud of gas
that would form the Milky Way experienced dissipational collapse in one smooth, adiabatic process. This is effective at forming the disk,
with only a tiny bit of star formation occurring during the collapse phase to provide the stars of the ancient, metal-poor stellar halo.
In contrast, the Galaxy could have been built up by the merger of smaller protogalactic fragments, 
each with their own life as smaller galaxies prior to merging. The latter is more natural
to the emergence of structure from the initial conditions observed in the CMB, where small lumps condense more readily than large ones.
Indeed, this effectively forms the basis of the modern picture of hierarchical galaxy formation \citep{hierarchicalGF}.

Hierarchical galaxy formation is effective at forming bulges and pressure-supported ETGs, but is anathema to the formation of orderly disks.
Dynamically cold disks are fragile and prefer to be left alone: the high rate of merging in the hierarchical \LCDM\ model tends to 
destroy the dynamically cold state in which most spirals are observed to exist \citep{TothO,Abadi,Peebles2020}. 
Consequently, there have been some rather different ideas about galaxy formation: 
if one starts from the initial conditions imposed by the CMB, hierarchical galaxy formation is inevitable.
If instead one works backwards from the observed state of galaxy disks, the smooth settling of gaseous disks in 
relatively isolated monoliths seems more plausible. 

In addition to different theoretical notions, our picture of the galaxy population was woefully incomplete. 
An influential study by \citet{F70} found that 28 of three dozen spirals shared very nearly the same central surface brightness.
This was generalized into a belief that all spirals had the same (high) surface brightness, and came to be known as Freeman's Law.
Ultimately this proved to be a selection effect, as pointed out early by \citet{Disney} and \citet{AllenShu}. However,
it was not until much later \citep{MBS95} that this became widely recognized. In the mean time, the prevailing assumption
was that Freeman's Law held true \citep[e.g.,][]{vdKspin} and all spirals had practically the same surface brightness.
In particular, it was the central surface brightness of the disk component of spiral galaxies that was thought to be universal,
while bulges and ETGs varied in surface brightness. Variation in the disk component of LTGs was thought to be restricted to
variations in size, which led to variations in luminosity at fixed surface brightness. 

Consequently, most theoretical effort was concentrated on the bright objects in the high-mass ($M_* > 10^{10}\;\Msun$) clump in Fig.\ \ref{fig:sizemass}.
Some low mass dwarf galaxies were known to exist, but were considered to be insignificant because they contained little mass. 
Low surface brightness galaxies violated Freeman's Law, so were widely presumed not to exist, or be at most a rare curiosity \citep{BosmaFreeman}.
A happy consequence of this unfortunate state of affairs was that as observations of diffuse LSB galaxies were made,
they forced then-current ideas about galaxy formation into a regime that they had not anticipated, and which many could not accommodate.

\begin{figure}[t]
\begin{center}
\includegraphics[width=1.0\textwidth]{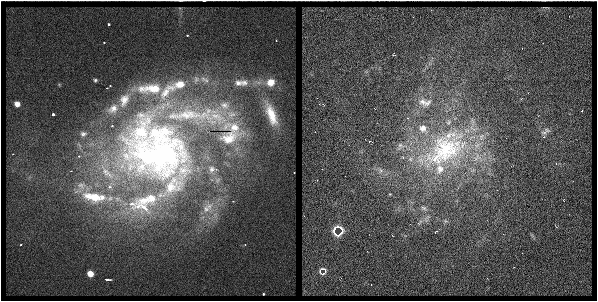}
\end{center}
\caption{High and low surface brightness galaxies. 
NGC 7757 (left) and UGC 1230 (right) are examples of high and low surface brightness galaxies, respectively.
These galaxies are about the same distance away and span roughly the same physical diameter. 
The chief difference is in the surface brightness, which follows from the separation between stars \citep{LSBmorphology}.
Note that the intensity scale of these images is not identical; the contrast has been increased for the LSB galaxy
so that it appears as more than a smudge. 
}
\label{fig:HSBLSB}
\end{figure}

The similarity and difference between high surface brightness (HSB) and LSB galaxies is illustrated by Fig.\ \ref{fig:HSBLSB}.
Both are rotationally supported, late type disk galaxies. Both show spiral structure, though it is more prominent in the HSB.
More importantly, both systems are of comparable linear diameter. 
They exist roughly at opposite ends of a horizontal line in Fig.\ \ref{fig:sizemass}. 
Their differing stellar masses stem from the surface density of their stars rather than their linear extent --- exactly the opposite
of what had been inferred from Freeman's Law. 
Any model of galaxy formation and evolution must account for the distribution of size (or surface brightness) at a given mass
as well as the number density of galaxies as a function of mass. Both aspects of the galaxy population remain problematic to this day. 

\subsection{Two Hypotheses: Spin and Density}
\label{sec:DDSH}

Here I discuss two basic hypotheses for the distribution of disk galaxy size at a given mass.
These broad categories I label \textbf{SH} (Same Halo) and \textbf{DD} (Density begets Density) following \citet{MdB98a}. 
In both cases, galaxies of a given baryonic mass are assumed to reside in dark matter halos of a corresponding total mass. 
Hence, at a given halo mass, the baryonic mass is the same, and variations in galaxy size follow from one of two basic effects:
\begin{itemize}
\item \textbf{SH:} variations in size follow from variations in the spin of the parent dark matter halo. 
\item \textbf{DD:} variations in surface brightness follow from variations in the density of the dark matter halo.
\end{itemize}
Recall that at a given luminosity, size and surface brightness are not independent, so variation in one corresponds to variation in the other.
Consequently, we have two distinct ideas for why galaxies of the same mass vary in size. In SH, the halo may have the same 
density profile $\rho(r)$, and it is only variations in angular momentum that dictate variations in the disk size. 
In DD, variations in the surface brightness of the luminous disk are reflections of variations in the density profile $\rho(r)$
of the dark matter halo. In principle, one could have a combination of both effects, 
but we will keep them separate for this discussion, and note that mixing them defeats the virtues of each without curing their ills.

The SH hypothesis traces back to at least \citet{FE_AM}. 
The notion is simple: variations in the size of disks correspond to variations in the angular momentum of their host dark matter halos. 
The mass destined to become a dark matter halo initially expands with the rest of the universe, reaching some maximum radius
before collapsing to form a gravitationally bound object. At the point of maximum expansion, the nascent dark matter halos
torque one another, inducing a small but non-zero net spin in each, quantified by the dimensionless 
spin parameter $\lambda$ \citep{spinparam}. One then imagines that as a disk forms within a dark matter halo,
it collapses until it is centrifugally supported: $\lambda \rightarrow 1$ from some initially small value 
\citep[typically $\lambda \approx 0.05$,][with some modest distribution about this median value]{spindistribution}. 
The spin parameter thus determines the collapse factor and the extent of the disk: 
low spin halos harbor compact, high surface brightness disks while high spin halos produce extended, low surface brightness disks.

The distribution of primordial spins is fairly narrow, and does not correlate with environment \citep{spindistribution}.
The narrow distribution was invoked as an explanation for Freeman's Law: the small variation in spins from halo to halo
resulted in a narrow distribution of disk central surface brightness \citep{vdKspin}. This association, while apparently natural, proved
to be incorrect: when one goes through the mathematics to transform spin into scale length, even a narrow distribution of initial spins 
predicts a broad distribution in surface brightness \citep{DSS97,MdB98a}. Indeed, it predicts too broad a distribution:
to prevent the formation of galaxies much higher in surface brightness than observed, one must invoke a 
stability criterion \citep{DSS97,MdB98a} that precludes the existence of very high surface brightness disks.
While it is physically quite reasonable that such a criterion should exist \citep{OP73}, the observed surface density threshold 
does not emerge naturally, and must be inserted by hand. It is an auxiliary hypothesis invoked to preserve SH.
Once done, size variations and the trend of average size with mass work out in reasonable quantitative detail \citep[e.g.,][]{MMW98}. 

Angular momentum conservation must hold for an isolated galaxy, but the assumption made in SH is stronger: 
baryons conserve their share of the angular momentum independently of the dark matter. 
It is considered a virtue that this simple assumption leads to disk sizes that are about right. 
However, this assumption is not well justified. 
Baryons and dark matter are free to exchange angular momentum with each other, and are seen to do so in simulations that track both 
components \citep[e.g.,][]{LZSangmom,BBPBGangmom,Combesangmom}. There is no guarantee that this exchange is
equitable, and in general it is not: as baryons collapse to form a small galaxy within a large dark matter halo, they tend to
lose angular momentum to the dark matter. This is a one-way street that runs in the wrong direction, with the final destination
uncomfortably invisible with most of the angular momentum sequestered in the unobservable dark matter. Worse still, 
if we impose rigorous angular momentum conservation among the baryons, the result is a disk with a completely unrealistic
surface density profile \citep{vdBangmom}. It then becomes necessary to pick and choose which baryons manage to assemble
into the disk and which are expelled or otherwise excluded, thereby solving one problem by creating another.

Early work on LSB disk galaxies led to a rather different picture.
Compared to the previously known population of HSB galaxies around which our theories had been built,
the LSB galaxy population has a younger mean stellar age \citep{MB94,dBvdH98}, a lower content of heavy elements \citep{M94}, 
and a systematically higher gas fraction \citep{MdB97,gasrich}. 
These properties suggested that LSB galaxies evolve more gradually
than their higher surface brightness brethren: they convert their gas into stars over a much longer timescale \citep{SFMS}.
The obvious culprit for this difference is surface density: lower surface brightness galaxies have less gravity,
hence less ability to gather their diffuse interstellar medium into dense clumps that could form stars \citep{MihosCO,deBlokCO}. 
It seemed reasonable to ascribe the low surface density of the baryons to a correspondingly low density of their parent dark matter halos.

One way to think about a region in the early universe that will eventually collapse to form a galaxy is as a so-called top-hat over-density.
The mass density $\Omega_m \rightarrow 1$ at early times, irrespective of its current value, so a spherical region (the top-hat) that
is somewhat over-dense early on may locally exceed the critical density. We may then consider this finite region as its own little closed
universe, and follow its evolution with the Friedmann equations with $\Omega > 1$. The top-hat will initially expand along with the rest of
the universe, but will eventually reach a maximum radius and recollapse. When that happens depends on the density.
The greater the over-density, the sooner the top-hat will recollapse. Conversely, a lesser over-density will take longer to reach 
maximum expansion before recollapsing. 

Everything about LSB galaxies suggested that they were lower density, late-forming systems.
It therefore seemed quite natural to imagine a distribution of over-densities and corresponding collapse times for top-hats of
similar mass, and to associate LSB galaxy with the lesser over-densities \citep{DS,mythesis}. 
More recently, some essential aspects of this idea have been revived under the monicker of ``assembly bias'' \citep[e.g.][]{assemblybias}.

The work that informed the DD hypothesis was based largely on photometric and spectroscopic observations of LSB galaxies:
their size and surface brightness, color, chemical abundance, and gas content. DD made two obvious predictions that had
not yet been tested at that juncture. First, late-forming halos should reside preferentially in low density environments. 
This is a generic consequence of Gaussian initial conditions: big peaks defined
on small (e.g., galaxy) scales are more likely to be found in big peaks defined on large (e.g., cluster) scales, and vice-versa.
Second, the density of the dark matter halo of an LSB galaxy should be lower than that of an equal mass halo containing and
HSB galaxy. This predicts a clear signature in their rotation speeds, which should be lower for lower density.

The prediction for the spatial distribution of LSB galaxies was tested by \citet{smallscaleLSB} and \citet{MMBsmallscale}. 
The test showed the expected effect: LSB galaxies were less strongly clustered than HSB galaxies. 
They are clustered: both galaxy populations follow the same large scale 
structure, but HSB galaxies adhere more strongly to it. In terms of the correlation function, the LSB sample available at the time had
about half the amplitude $r_0$ as comparison HSB samples \citep{MMBsmallscale}. The effect was even more pronounced on the smallest
scales \citep[$< 2$ Mpc:][]{smallscaleLSB}, leading \citet{MMBsmallscale} to construct a model that successfully explained both small
and large scale aspects of the spatial distribution of LSB galaxies simply by associating them with dark matter halos that lacked 
close interactions with other halos. This was strong corroboration of the DD hypothesis.

\begin{figure}[t]
\begin{center}
\includegraphics[width=1.0\textwidth]{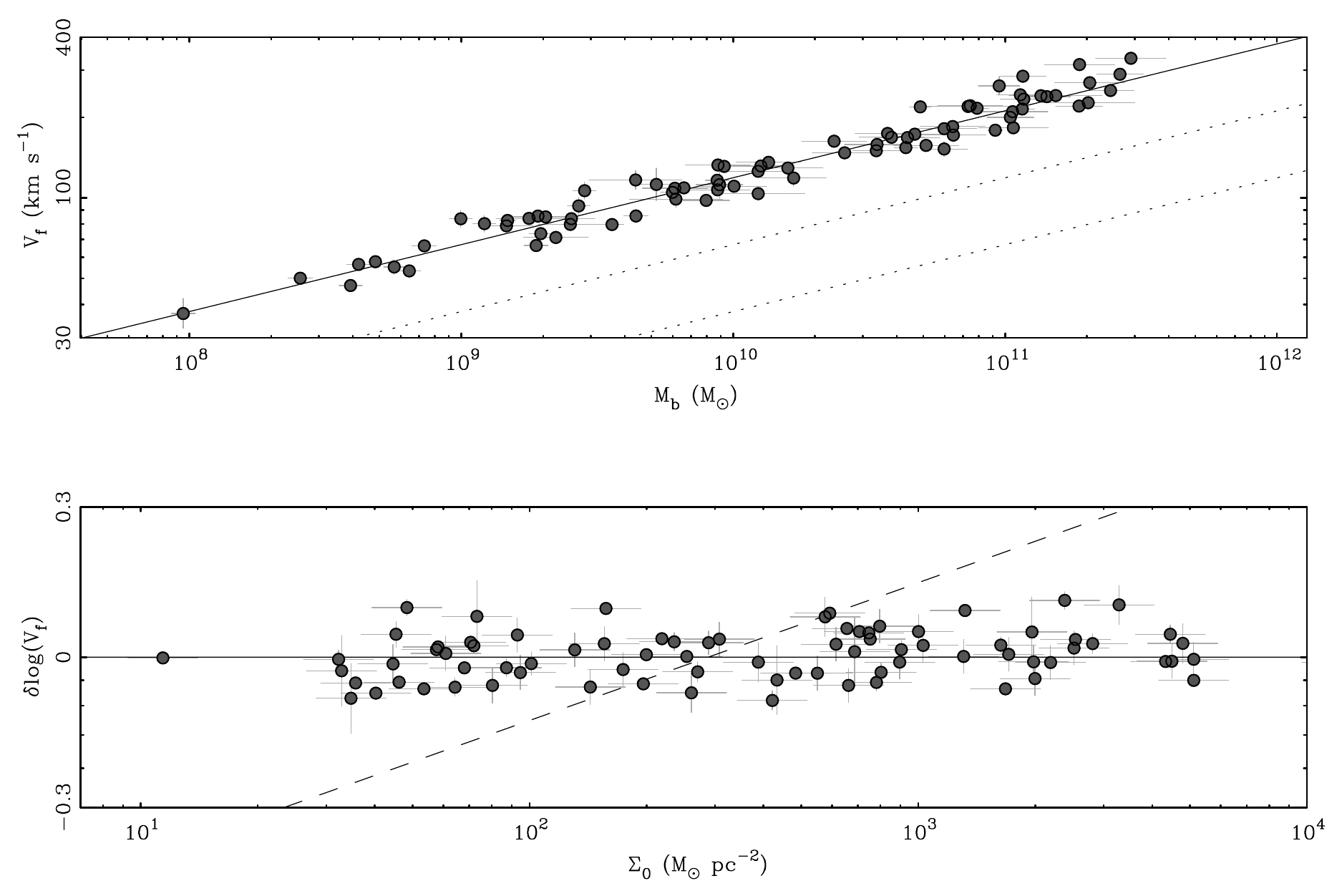}
\end{center}
\caption{The Baryonic Tully-Fisher relation and residuals. The top panel shows the flat rotation velocity of galaxies in the SPARC database \citep{SPARC}
as a function of the baryonic mass (stars plus gas). The sample is restricted to those objects for which both quantities are measured 
to better than 20\% accuracy. The bottom panel shows velocity residuals around the solid line in the top panel as a function of the central
surface density of the stellar disks. Variations in the stellar surface density predict variations in velocity along the dashed line. These would
translate to shifts illustrated by the dotted lines in the top panel, with each dotted line representing a shift of a factor of ten in surface density.
The predicted dependence on surface density is not observed \citep{zwaanTF,sprayTF,MdB98a,CR}.
}
\label{fig:BTFRandResid}
\end{figure}

One way to test the prediction of DD that LSB galaxies should rotate more slowly than HSB galaxies was to use the
Tully-Fisher relation \citep{TForig} as a point of reference.  
Originally identified as an empirical relation between optical luminosity and the observed line-width of single-dish 21cm observations,
more fundamentally it turns out to be a relation between the baryonic mass of a galaxy (stars plus gas) and its flat rotation speed
the Baryonic Tully-Fisher relation \citep[BTFR:][]{btforig}. This relation is a simple power law of the form
\begin{equation}
M_b = A V_f^4
\end{equation} 
with $A \approx 50\;\Msun\,\mathrm{km}^{-4}\,\mathrm{s}^4$ \citep{M05}.

\citet{AaronsonTF} provided a straightforward interpretation for a relation of this form.
A test particle orbiting a mass $M$ at a distance $R$ will have a circular speed $V$
\begin{equation}
V^2 = GM/R
\end{equation} 
where $G$ is Newton's constant. If we square this, a relation like the Tully-Fisher relation follows:
\begin{equation}
V^4 = (GM/R)^2 \propto M \Sigma
\label{eq:Sigma}
\end{equation} 
where we have introduced the surface mass density $\Sigma = M/R^2$. 
The Tully-Fisher relation $M \propto V^4$ is recovered if $\Sigma$ is constant, exactly as expected from Freeman's Law \citep{F70}.

LSB galaxies, by definition, have central surface brightnesses (and corresponding stellar surface densities $\Sigma_0$) that are 
less than the Freeman value. Consequently, DD predicts, through equation \ref{eq:Sigma}, 
that LSB galaxies should shift systematically off the Tully-Fisher relation: lower $\Sigma$ means lower velocity.
The predicted effect is not 
subtle\footnote{If I had believed that we could get away with a subtle shift, I would have patched up my hypothesis accordingly.
Since it was not possible, I rejected my own hypothesis.} 
(Fig.\ \ref{fig:BTFRandResid}).
For the range of surface brightness that had become available, the predicted shift should have stood out like the proverbial sore thumb.
It did not \citep{zwaanTF,sprayTF,hoffmanTF,MdB98a}. 
This had an immediate impact on galaxy formation theory: 
compare \citet[who predict a shift in Tully-Fisher with surface brightness]{Dalc95} with \citet[who do not]{DSS97}.

\begin{figure}[t]
\begin{center}
\includegraphics[width=1.0\textwidth]{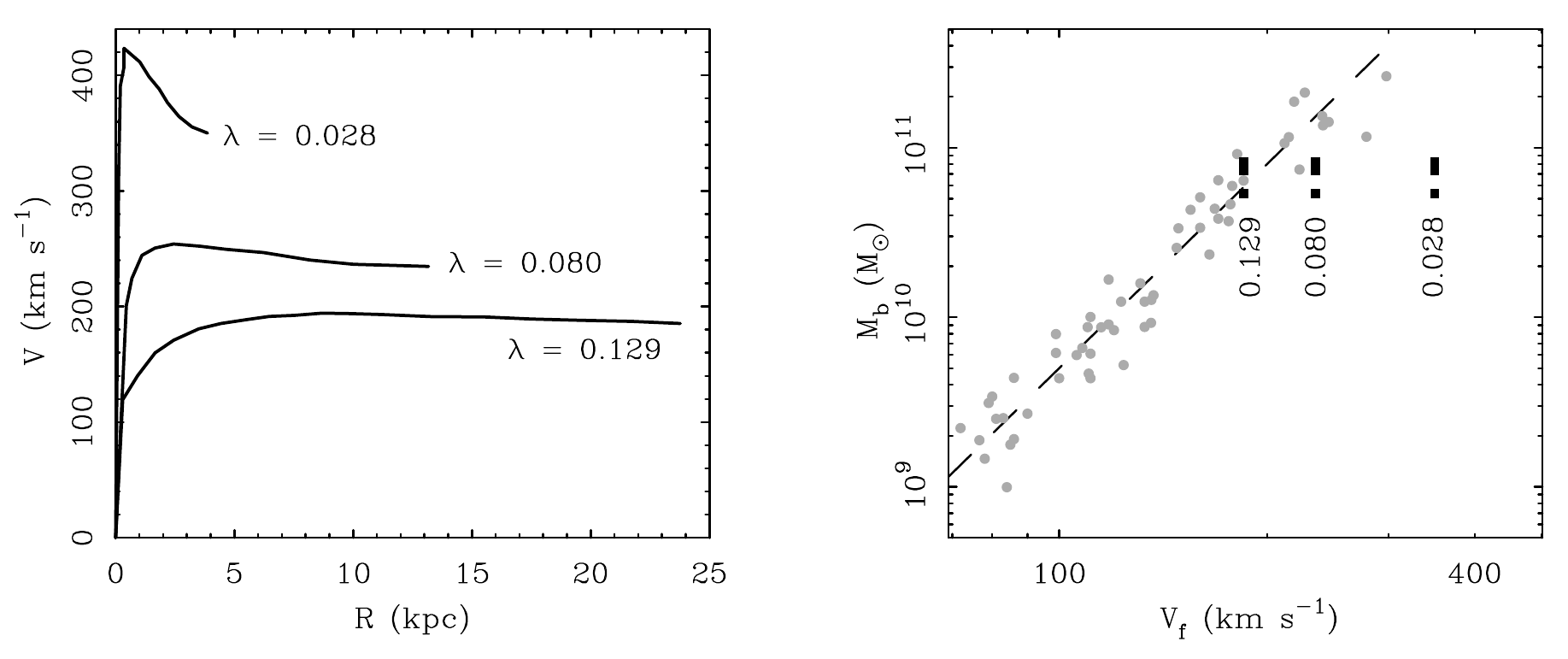}
\end{center}
\caption{Model galaxy rotation curves and the Tully-Fisher relation.
Rotation curves (left panel) for model galaxies of the same mass but different spin parameters $\lambda$ 
from \citet[see his Fig.\ 3]{vdB01}. Models with lower spin have more compact stellar disks that contribute more to the
rotation curve ($V^2 = GM/R$; $R$ being smaller for the same $M$). These models are shown as square points on the Baryonic Tully-Fisher
relation (right) along with data for real galaxies \citep[grey circles:][]{SPARCTF} and a fit thereto (dashed line). 
Differences in the cooling history result in modest variation in the baryonic mass at fixed halo mass as reflected in the vertical scatter 
of the models. This is within the scatter of the data, but variation due to the spin parameter is not.
}
\label{fig:vdBVRTF}
\end{figure}

Instead of the systematic variation of velocity with surface brightness expected at fixed mass, there was none. 
Indeed, there is no hint of a second parameter dependence. The relation is incredibly tight by the standards
of extragalactic astronomy \citep{SPARCTF}: baryonic mass and the flat rotation speed are practically interchangeable. 

The above derivation is overly simplistic. The radius at which we should make a measurement is ill-defined, and the surface density
is dynamical: it includes both stars and dark matter. Moreover, galaxies are not spherical cows: 
one needs to solve the Poisson equation for the observed disk geometry of LTGs, and account for the varying radial contributions of
luminous and dark matter. While this can be made to sound intimidating, the numerical computations are straightforward and 
rigorous \citep[e.g.,][]{Casertano,BBS,SPARC}. It still boils down to the same sort of relation (modulo geometrical factors of order unity),
but with two mass distributions: one for the baryons $M_b(R)$, and one for the dark matter $M_{DM}(R)$. 
Though the dark matter is more massive, it is also more extended. Consequently, both components can contribute non-negligibly to
the rotation over the observed range of radii:
\begin{equation}
V^2(R) = \frac{GM}{R} = G\left(\frac{M_b(R)}{R}+\frac{M_{DM}(R)}{R}\right),
\label{eq:diskhalo}
\end{equation} 
where for clarity we have omitted\footnote{Strictly speaking, eq.\ \ref{eq:diskhalo} only holds for spherical mass distributions. 
I make this simplification here to emphasize the fact that both mass and radius matter.} geometrical factors.
The only absolute requirement is that the baryonic contribution should begin to 
decline once the majority of baryonic mass is encompassed.
It is when rotation curves persist in remaining flat past this point that we infer the need for dark matter.

A recurrent problem in testing galaxy formation theories is that they seldom make ironclad predictions; 
I attempt a brief summary in Table \ref{tab:DDSH}.
SH represents a broad class of theories with many variants. 
By construction, the dark matter halos of galaxies of similar stellar mass are similar. \textit{If} we associate
the flat rotation velocity with halo mass, then galaxies of the same mass have the same circular velocity, 
and the problem posed by Tully-Fisher is automatically satisfied. 

While it is common to associate
the flat rotation speed with the dark matter halo, this is a half-truth: 
the observed velocity is a combination of baryonic and dark components (eq.\ \ref{eq:diskhalo}). 
It is thus a rather curious coincidence that rotation curves are as flat as they are: 
the Keplerian decline of the baryonic contribution must be precisely balanced by an increasing contribution from
the dark matter halo. This fine-tuning problem was dubbed the ``disk-halo conspiracy'' \citep{BC85,vAS1986}.
The solution offered for the disk-halo conspiracy was that the formation of the baryonic disk has an effect on the
distribution of the dark matter. As the disk settles, the dark matter halo respond through a process
commonly referred to as adiabatic compression that brings
the peak velocities of disk and dark components into alignment \citep{Blumenthal}. 
Some rearrangement of the dark matter halo in response to the change of the gravitational potential caused by the settling
of the disk is inevitable, so this seemed a plausible explanation. 

The observation that LSB galaxies obey the Tully-Fisher relation greatly compounds the fine-tuning \citep{zwaanTF,MdB98a}.
The amount of adiabatic compression depends on the surface density of stars \citep{adiabat}:
HSB galaxies experience greater compression than LSB galaxies. 
This should enhance the predicted shift between the two in Tully-Fisher.
Instead, the amplitude of the flat rotation speed remains unperturbed.

\begin{table}[t]
\caption{Predictions of DD and SH for LSB Galaxies}
 \label{tab:DDSH}
\begin{center}
\begin{tabular}{l c c}
\textbf{Observation} & \textbf{DD} & \textbf{SH} \\
Evolutionary rate & + & + \\
Size distribution & + & + \\
Clustering & + & X \\
Tully-Fisher relation & X & ? \\
Central density relation & + & X
\end{tabular}
\end{center}
\end{table}

The generic failings of dark matter models was discussed at length by \citet{MdB98a}. 
The same problems have been encountered by others. For example, Fig.\ \ref{fig:vdBVRTF} shows model galaxies formed in a dark matter halo with identical
total mass and density profile but with different spin parameters \citep{vdB01}. Variations in the assembly and cooling history were also considered,
but these make little difference and are not relevant here. The point is that smaller (larger) spin parameters lead to more (less) compact disks
that contribute more (less) to the total rotation, exactly as anticipated from variations in the term $M_b/R$ in equation \ref{eq:diskhalo}.
The nominal variation is readily detectable, and stands out prominently in the Tully-Fisher diagram (Fig.\ \ref{fig:vdBVRTF}).
This is exactly the same fine-tuning problem that was pointed out by \citet{zwaanTF} and \citet{MdB98a}.

\begin{figure}[t]
\begin{center}
\includegraphics[width=1.0\textwidth]{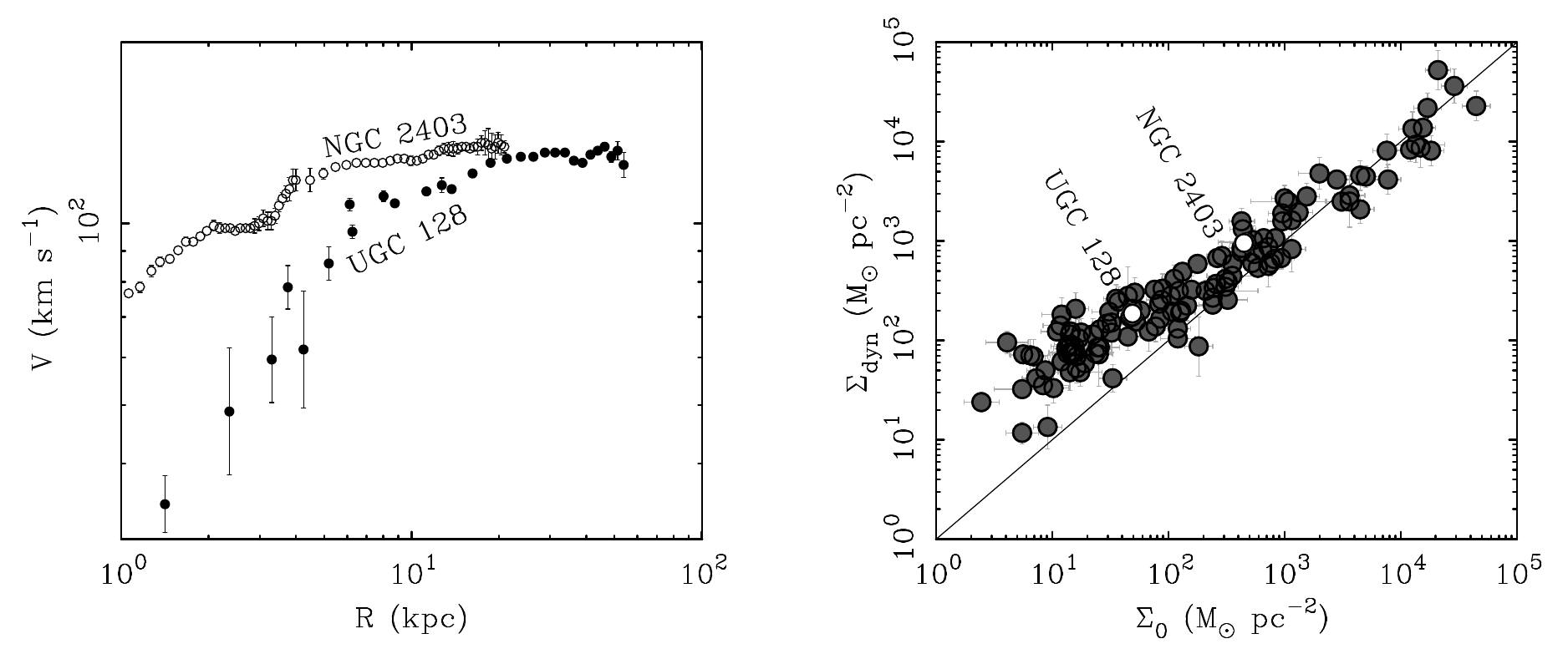}
\end{center}
\caption{Rotation curve shapes and surface density. 
The left panel shows the rotation curves of two galaxies, one HSB (NGC 2403, open circles) and 
one LSB (UGC 128, filled circles) \citep{dBM96,VdB1999,KdN08}. 
These galaxies have very nearly the same baryonic mass ($\sim 10^{10}\;\Msun$), 
and asymptote to approximately the same flat rotation speed ($\sim 130\;\kms$).
Consequently, they are indistinguishable in the Tully-Fisher plane (Fig.\ \ref{fig:BTFRandResid}). 
However, the inner shapes of the rotation curves are readily distinguishable:
the HSB galaxy has a steeply rising rotation curve while the LSB galaxy has a more gradual rise.
This is a general phenomenon, as illustrated by the central density relation \citep[right panel:][]{CDR}
where each point is one galaxy; NGC 2403 and UGC 128 are highlighted as open points.
The central dynamical mass surface density ($\Sigma_{\mathrm{dyn}}$) measured by 
the rate of rise of the rotation curve \citep{ToomreCDR} correlates with 
the central surface density of the stars ($\Sigma_0$) measured by their surface brightness. 
The line shows 1:1 correspondence: no dark matter is required near the centers of HSB galaxies.
The need for dark matter appears below $1000\;\Msun\,\mathrm{pc}^{-2}$ and
grows systematically greater to lower surface brightness. 
This is the origin of the statement that LSB galaxies are dark matter dominated.
   \label{fig:RCandCDR}}
\end{figure}

What I describe as a fine-tuning problem is not portrayed as such by \citet{vdB00} and \citet{vdBDalc00}, who argued
that the data could be readily accommodated in the dark matter picture. The difference is between accommodating
the data once known, and predicting it \textit{a priori}. The dark matter picture is extraordinarily flexible: one is free to distribute
the dark matter as needed to fit any data that evinces a non-negative mass discrepancy, even data that are wrong \citep{dBM98}. 
It is another matter entirely to construct a realistic model \textit{a priori}; in my experience it is quite easy to construct models
with plausible-seeming parameters that bear little resemblance to real galaxies (e.g., the low-spin case in Fig.\ \ref{fig:vdBVRTF}).
A similar conundrum is encountered when constructing models that can explain the long tidal tails observed in merging and
interacting galaxies: models with realistic rotation curves do not produce realistic tidal tails, and vice-versa \citep{DMH}.
The data occupy a very narrow sliver of the enormous volume of
parameter space available to dark matter models, a situation that seems rather contrived. 

Both DD and SH predict residuals from Tully-Fisher that are not observed. I consider this to be an unrecoverable failure for 
DD, which was \textit{my} hypothesis \citep{mythesis}, so I worked hard to salvage it. I could not.
For SH, Tully-Fisher might be recovered in the limit of dark matter domination, which requires further consideration. 

\subsection{Squeezing the Toothpaste Tube}

Our efforts to evade one fine-tuning problem often lead to another.
This has been my general experience in many efforts to construct viable dark matter models.
It is like squeezing a tube of toothpaste: every time we smooth out the problems in one part of the tube,
we simply squeeze them into a different part. There are many published claims to solve this problem or that, 
but they frequently fail to acknowledge (or notice) that the purported solution to one problem creates another.

One example is provided by \citet{CR}. They invoke dark matter domination to explain the lack of residuals in 
theTully-Fisher relation. In this limit, $M_b/R \ll M_{DM}/R$ and the baryons leave no mark on the rotation curve.
This can reconcile the model with the Tully-Fisher relation, but it makes a strong prediction.
It is not just the flat rotation speed that is the same for galaxies of the same mass, but the entirety of the rotation curve, $V(R)$
at all radii. The stars are
just just convenient tracers of the dark matter halo in this limit; the dynamics are entirely dominated by the dark matter.
The hypothesized solution fixes the problem that is addressed, but creates another problem that is not addressed,
in this case the observed variation in rotation curve shape.  

The limit of complete dark matter domination is not consistent with the shapes of rotation curves. 
Galaxies of the same baryonic mass have the same flat outer velocity (Tully-Fisher), 
but the shapes of their rotation curves vary systematically with surface 
brightness \citep{dBM96,TVbimodal,MdB98a,MdB98b,Swaters09,Swaters12,LelliVRgrad,CDR}.
High surface brightness galaxies have steeply rising rotation curves while LSB galaxies have slowly rising 
rotation curves (Fig.\ \ref{fig:RCandCDR}).
This systematic dependence of the inner rotation curve shape on the baryon distribution excludes the SH hypothesis 
in the limit of dark matter domination: the distribution of the baryons clearly has an impact on the dynamics.

A more recent example of this toothpaste tube problem for SH-type models is provided by the EAGLE simulations \citep{EAGLE}.
These are claimed \citep{Ludlow2017} to explain one aspect of the observations, the radial acceleration relation \citep{RAR},
but fail to explain another, the central density relation \citep{CDR} seen in Fig.\ \ref{fig:RCandCDR}.
This was called the `diversity' problem by \citet{diversity}, who note that the rotation velocity at a specific, small radius (2 kpc) 
varies considerably from galaxy to galaxy observationally (Fig.\ \ref{fig:RCandCDR}), 
while simulated galaxies show essentially no variation, with only a small amount of scatter.
This diversity problem is exactly the same problem that was pointed out 
before [compare Fig.\ 5 of \citet{diversity} to Fig.\ 14 of \citet{MdB98a}].

\begin{figure}[t]
\begin{center}
\includegraphics[width=1.0\textwidth]{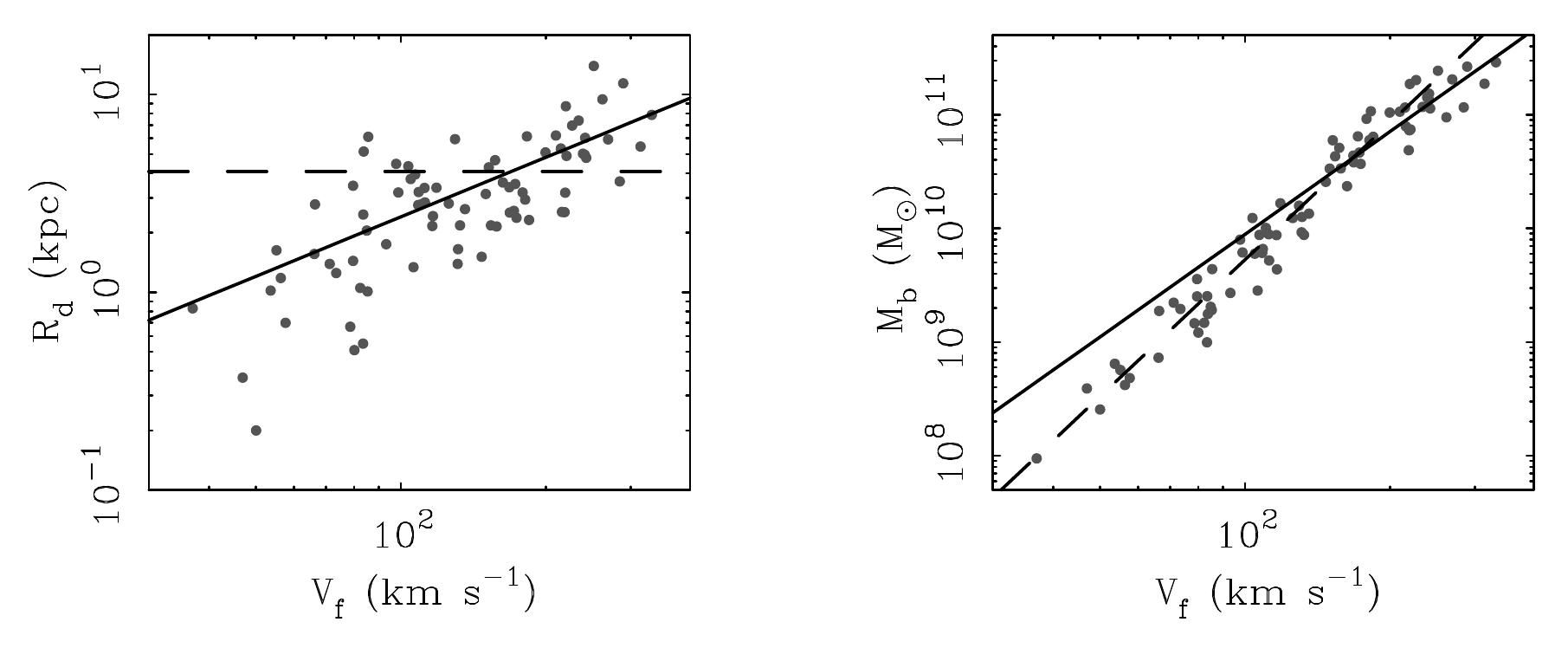}
\end{center}
\caption{Galaxy size (as measured by the exponential disk scale length, left) and mass (right) as a function of rotation velocity. 
The latter is the Baryonic Tully-Fisher relation; the data are the same as in Fig.\ \ref{fig:BTFRandResid}.
The solid lines are \citet{MMW98} models with constant $m_d$ (their equations 12 and 16). 
This is in reasonable agreement with the size-speed relation but not the BTFR. 
The latter may be fit by adopting a variable $m_d \propto V_f$ (dashed lines), but this ruins agreement with the size-speed relation.
This is typical of dark matter models in which fixing one thing breaks another.
   \label{fig:RdMbVf}}
\end{figure}

There is no single, universally accepted standard galaxy formation model, but a common touchstone is provided by \citet{MMW98}.
Their base model has a constant ratio of luminous to dark mass $m_d$  [their assumption (i)], which 
provides a reasonable description of the sizes of galaxies as a function of mass or rotation speed (Fig.\ \ref{fig:RdMbVf}). However, this model 
predicts the wrong slope (3 rather than 4) for the Tully-Fisher relation. This is easily remedied by 
by making the luminous mass fraction proportional to the rotation speed ($m_d \propto V_f$), which then provides an adequate fit to
the Tully-Fisher\footnote{The normalization of the Tully-Fisher relation depends on an appropriate but arbitrary choice of $m_d$.}
relation. This has the undesirable effect of destroying the consistency of the size-mass relation. We can have
one or the other, but not both.

This failure of the \citet{MMW98} model provides another example of the toothpaste tube problem. By fixing one problem, we create another.
The only way forward is to consider more complex models with additional degrees of freedom.

\subsection{Feedback}

It has become conventional to invoke `feedback' to address the various problems that afflict galaxy formation theory.
Feedback is not a single process, but rather a family of distinct processes.
The common feature of different forms of feedback is the deposition of energy from compact sources into the surrounding gas 
of the interstellar medium. This can, at least in principle,  
heat gas and drive large-scale winds, either preventing gas from cooling and forming too many stars, or ejecting
it from a galaxy outright. This in turn might affect the distribution of dark matter, thought the effect is weak:
one must move a lot of baryons for their gravity to impact the dark matter distribution.  

There are many kinds of feedback, and many devils in the details. 
Massive, short-lived stars produce copious amounts of ultraviolet radiation that heats and
ionizes the surrounding gas and erodes interstellar dust. 
These stars also produce strong winds through much of their short ($\sim 10$ Myr) lives,
and ultimately explode as Type II supernovae. These three mechanisms each act in a distinct way on different time scales.
That's just the feedback associated with massive stars; there are many other mechanisms 
(e.g., Type Ia supernovae are distinct from Type II supernovae, and Active Galactic Nuclei are a completely different beast entirely).
The situation is extremely complicated. While the various forms of stellar feedback are readily apparent on the small scales of stars,
it is far from obvious that they have the desired impact on the much larger scales of entire galaxies. 

For any one kind of feedback, there can be many substantially different implementations in galaxy formation simulations.
Independent numerical codes do not generally return compatible results for identical initial conditions \citep{Aquila}:
there is no consensus on how feedback works. 
Among the many different computational implementations of feedback, at most one can be correct.

Most galaxy formation codes do not resolve the scale of single stars where stellar feedback occurs.
They rely on some empirically calibrated, analytic approximation to model this `sub-grid physics' --- 
which is to say, they don't simulate feedback at all. Rather, they simulate the accumulation of gas in one resolution
element, then follow some prescription for what happens inside that unresolved box.
This provides ample opportunity for disputes over the implementation and effects of feedback.
For example, feedback is often cited as a way to address the cusp-core problem --- or not, depending on the
implementation \citep[e.g.,][]{Governato2012,DC14,Madau14,Read_cNFW,Bose19,nostinkingcores}.
High resolution simulations \citep{hirezJBH} indicate that the gas of the interstellar medium is less affected
by feedback effects than assumed by typical sub-grid prescriptions: most of the energy is funneled through the lowest
density gas --- the course of least resistance --- and is lost to the intergalactic medium without much impacting the
galaxy in which it originates. 

From the perspective of the philosophy of science, feedback is an auxiliary hypothesis invoked to patch up theories of
galaxy formation. Indeed, since there are many distinct flavors of feedback that are invoked to carry out a variety of
different tasks, feedback is really a suite of auxiliary hypotheses. There is, at present, widespread agreement
that feedback is necessary, but little consensus on how it works, or whether it is possible for it to solve all of the problems for which it is invoked.

The problem with the current state of galaxy formation theory is not specific to any particular feedback scheme.
What is troubling is that complex processes are invoked to explain simple phenomena. 
There are more free parameters than constraints, so 
there is no doubt that models can be built to match the data --- even data that bear little resemblance to
the obvious predictions of the paradigm \citep{MMW98,MdB98a}. 
The concern is not that \LCDM\ galaxy formation models do not explain the data; it is that they can't not. 

\section{Modified Newtonian Dynamics}
\label{sec:MOND}

There is one and only one theory that predicted in advance the observations described above:
the Modified Newtonian Dynamics (MOND) introduced by \citet{milgrom83b,milgrom83c,milgrom83a}.
MOND is an extension of Newtonian theory \citep{milgromtheseproc}. 
It is not a generally covariant theory, so is not, by itself, a complete replacement for General Relativity.
Nevertheless, it makes unique, testable predictions within its regime of applicability \citep{M2020}.

The basic idea of MOND is that the force law is modified at an acceleration scale, \azero.
For large accelerations, $g \gg \azero$, everything is normal and Newtonian: $g = g_N$, where $g_N$ is the acceleration
predicted by the observed luminous mass distribution obtained by solving the Poisson equation. 
At low accelerations, the effective acceleration tends towards the limit 
\begin{equation}
g \rightarrow \sqrt{\azero g_N} \;\mathrm{for}\; g \ll \azero
\label{eq:mond}
\end{equation}
\citep{milgrom83a,AQUAL}.
This limit is called the deep MOND regime in contrast to the Newtonian regime at high accelerations.
The two regimes are smoothly connected by an interpolation function $\mu(g/\azero)$ that is not specified \citep{milgrom83a}.

The motivation to make an acceleration-based modification is to explain flat rotation curves \citep{vera,bosma} that also gives
a steep Tully-Fisher relation similar to that which is observed \citep{AaronsonTF}. A test particle in a circular orbit around a point 
mass $M_p$ in the deep MOND regime (eq.\ \ref{eq:mond}) will experience a centripetal acceleration
\begin{equation}
\frac{V_c^2}{R} = \sqrt{\azero \frac{G M_p}{R^2}}.
\label{eq:centrip}
\end{equation}
Note that the term for the radius $R$ cancels out, so eq.\ \ref{eq:centrip} reduces to 
\begin{equation}
V_c^4 = \azero GM_p
\label{eq:mondTF}
\end{equation}
which the reader will recognize as the Baryonic Tully-Fisher relation 
\begin{equation}
M_b = A V_f^4
\label{eq:BTFR}
\end{equation}
with\footnote{The factor $\zeta$ is a geometric correction factor of order unity that accounts
for the fact that flattened mass distributions rotate faster than the equivalent spherical mass distribution \citep{BT}. 
The point-mass approximation made for illustration in eq.\ \ref{eq:centrip} neglects this 
geometric effect, which applies in any theory. For realistic disk galaxies, $\zeta \approx 0.8$ \citep{MdB98b,M05}.} 
$A = \zeta/(\azero G)$. 

This simple math explains the flatness of rotation curves. 
This is not a prediction; it was an input that motivated the theory, as it motivated dark matter. 
Unlike dark matter, in which rotation curves might rise or fall, 
the rotation curves of isolated galaxies must tend towards asymptotic flatness.

MOND also explains the Tully-Fisher relation. Indeed, there are several distinct aspects to this prediction. 
That the relation exists at all is a strong prediction. Fundamentally, the Baryonic Tully-Fisher Relation (BTFR) is a relation between the baryonic mass
of a galaxy and its flat rotation speed. There is no dark matter involved: $V_f$ is not a property of a dark matter halo, but of the galaxy itself.

One MOND prediction is the slope of the BTFR: the power law scaling $M \sim V^x$ has $x = 4$ exactly. 
While the infrared data of \citet{AaronsonTF}
suggested such a slope, the exact value was not well constrained at that time. It was not until later that Tully-Fisher was 
empirically recognized as a relation driven by baryonic mass \citep{btforig}, as anticipated by MOND. Moreover, the slope
is only four when a good measurement of the flat rotation velocity is available \citep{verhTF,M05,M12}; common proxies
like the line-width only crudely approximate the result and typically return shallower slopes \citep[e.g.,][]{zaritskyTF}, as do
samples of limited dynamic range \citep[e.g.,][]{pizagnoTF}. The latter are common in the literature: 
selection effects strongly favor bright galaxies, and the majority of published Tully-Fisher relations are dominated by
high mass galaxies ($M_* > 10^{10}\;\Msun$). Consequently, the behavior of the Baryonic Tully-Fisher relation
remains somewhat controversial to this day \citep[e.g.,][]{offthebtfr,superspirals}. This appears to be entirely
a matter of data quality \citep{IAU_review}. The slope of the relation is indistinguishable from 4 when a modicum of quality
control is imposed \citep{M05,stark,trach,M12,SPARCTF,JSH0}.
Indeed, only a slope of four successfully predicted the rotation speeds of low mass galaxies \citep{M11,LeoPdisc}.

Another aspect of the Tully-Fisher relation is its normalization. This is set by fundamental constants:
Newton's constant, $G$, and the acceleration scale of MOND, \azero. 
For $\zeta = 0.8$, $A = 50\;\Msun\,\mathrm{km}^{-4}\,\mathrm{s}^4$. 
However, there is no theory that predicts the value of \azero, which has to be set by the data. Moreover, this scale is 
distance-dependent, so the precise value of \azero\ varies with adjustments to the distance scale. For this reason, in part,
the initial estimate of $\azero = 2 \times 10^{-10}\;\mathrm{m}\,\mathrm{s}^{-2}$ of \citep{milgrom83b} was a bit high.
\citet{BBS} used the best data then available to obtain $\azero = 1.2 \times 10^{-10}\;\mathrm{m}\,\mathrm{s}^{-2}$.
The value of Milgrom's acceleration constant has not varied meaningfully since then \citep{SMmond,M11,LivRev,RAR,LiRAR}.
This is a consistency check, but not a genuine\footnote{One may reverse the argument to use the value of \azero\ to
predict the value of the Hubble constant. The value of \azero\ found by \citet{BBS} predicts
$H_0 \approx 75\;\mathrm{km}\,\mathrm{s}^{-1}\,\mathrm{Mpc}^{-1}$, as commonly found by
subsequent direct measurements \citep[e.g.,][]{HSTKP,Tully75,Riess19,Freedman19,JSH0}.} prediction.

An important consequence of MOND is that the Tully-Fisher relation is absolute: it should have no
dependence on size or surface brightness \citep{milgrom83b}. The mass of baryons is the only thing that sets the flat
amplitude of the rotation speed. It matters not at all how those baryons are distributed. MOND was the only theory to 
correctly predict this in advance of the observation \citep{MdB98b}.
The fine-tuning problem that we face conventionally is imposed by this otherwise unanticipated result.

The absolute nature of the Tully-Fisher relation in MOND further predicts that it has no physical residuals whatsoever.
That is to say, scatter around the relation can only be caused by observational errors and scatter in
the mass-to-light ratios of the stars. The latter is an irreducible unknown: we measure the luminosity produced
by the stars in a galaxy, but what we need to know is the mass of those stars. The conversion between them can never
be perfect, and inevitably introduces some scatter into the relation. Nevertheless, we can make our best effort to account
for known sources of scatter. Between scatter expected from observational uncertainties and that induced by variations
in the mass-to-light ratio, the best data are consistent with the prediction of zero intrinsic scatter \citep{M05,M12,SPARCTF,Lelli19}.
Of course, it is impossible to measure zero, but it is possible to set an upper limit on the intrinsic scatter that is very
tight by extragalactic standards \citep[$< 6\%$][]{Lelli19}. This leaves very little room for variations beyond the inevitable 
impact of the stellar mass-to-light ratio.
The scatter is no longer entirely accounted for when lower quality data are considered \citep{M12}, but this is  
expected in astronomy: lower quality data inevitably admit systematic uncertainties that are not readily accounted for in the
error budget. 

\citet{milgrom83b} made a number of other specific predictions. In MOND, the acceleration expected for kinematics
follows from the surface density of baryons. Consequently, low surface brightness means low acceleration. Interpreted
in terms of conventional dynamics, the prediction is that the ratio of dynamical mass to light, $M_{\mathrm{dyn}}/L$ 
should increase as surface brightness decreases. This happens both globally --- LSB galaxies appear to be more
dark matter dominated than HSB galaxies \citep[see Fig.\ 4(b) of][]{MdB98a}, and locally --- the need for dark matter
sets in at smaller radii in LSB galaxies than in HSB galaxies \citep[Fig.\ 3, 14 of][respectively]{MdB98b,LivRev}. 

One may also test this prediction by plotting the rotation curves of galaxies binned by surface brightness: acceleration should scale
with surface brightness. It does \citep[Fig.\ 4, 16 of][respectively]{MdB98b,LivRev}. This observation has been confirmed
by near-infrared data. The systematic variation of color coded surface brightness is already obvious with
optical data, as in Fig.\ 15 of \citet{LivRev}, but these suffer some scatter from variations in the stellar mass-to-light ratio.
These practically vanish with near-infrared data, which provide such a good tracer of the surface mass density of stars that
the equivalent plot is a near-perfect rainbow \citep[Fig.\ 3 of both][]{IAU_review,M2020}. The data strongly corroborate
the prediction of MOND that acceleration follows from baryonic surface density.

The central density relation \citep[Fig.\ \ref{fig:RCandCDR},][]{CDR} was also predicted by MOND \citep{milgromCDR}.
Both the shape and the amplitude of the correlation are correct. 
Moreover, the surface density $\Sigma_{\dagger}$ at which the data bend follows directly from the acceleration scale of 
MOND: $\azero = G \Sigma_{\dagger}$. This surface density also corresponds to the stability limit for 
disks \citep{MOND_Freeman,bradastab}.
The scale we had to insert by hand in dark matter models is a consequence of MOND.

\begin{figure}[t]
\begin{center}
\includegraphics[width=1.0\textwidth]{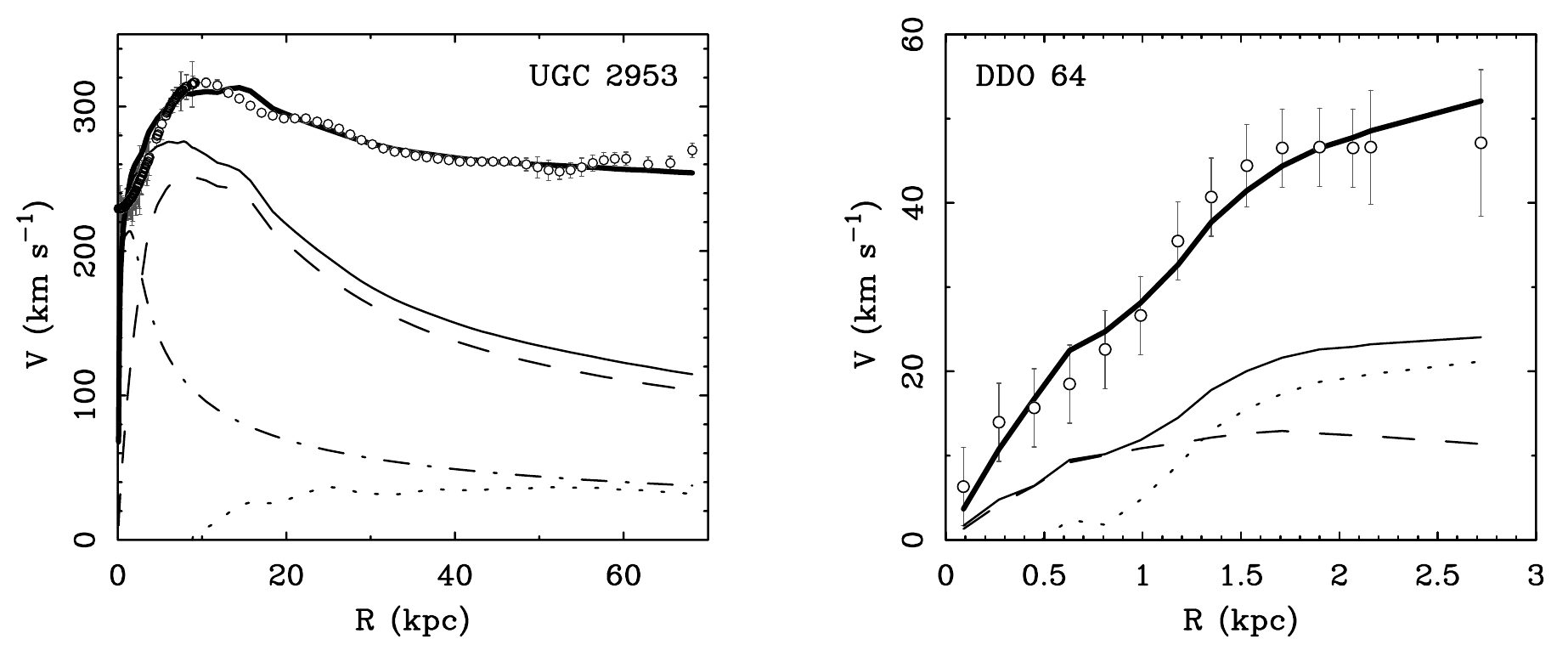}
\end{center}
\caption{Example rotation curve fits. MOND fits \citep[heavy solid lines:][]{LiRAR} to the rotation curves of a bright,
star-dominated galaxy (UGC 2953, left panel) and a faint, gas-dominated galaxy (DDO 64, right panel). 
The thin solid lines shows the Newtonian expectation, which is the sum of the
atomic gas (dotted lines), stellar disk (dashed lines), and stellar bulge (dash-dotted line; present only in UGC 2953). 
Note the different scales: UGC 2953 is approximately 400 times more massive than DDO 64.
   \label{fig:ugc2953ddo064}}
\end{figure}

Since MOND is a force law, the entirety of the rotation curve should follow from the baryonic mass distribution.  
The stellar mass-to-light ratio can modulate the amplitude of the stellar contribution to the rotation curve, but not its shape, which is specified by
the observed distribution of light. Consequently, there is rather limited freedom in fitting rotation curves.

Example fits are shown in Fig.\ \ref{fig:ugc2953ddo064}.
The procedure is to construct Newtonian mass models by numerically solving the Poisson equation 
to determine the gravitational potential that corresponds to the observed baryonic mass distribution.
Indeed, it is important to make a rigorous solution of the Poisson equation in order to capture details in the shape
of the mass distribution (e.g., the wiggles in Fig.\ \ref{fig:ugc2953ddo064}).
Common analytic approximations like the exponential disk assume these features out of existence.
Building proper mass models
involves separate observations for the stars, conducted at optical or near-infrared wavelengths,
and the gas of the interstellar medium, which is traced by radio wavelength observations.
It is sometimes necessary to consider separate mass-to-light ratios for the stellar bulge and disk components, 
as there can be astrophysical differences between these distinct stellar populations \citep{stellarpops}. 
This distinction applies in any theory.

The gravitational potential of each baryonic component is represented by the circular velocity
of a test particle in Fig.\ \ref{fig:ugc2953ddo064}. The amplitude of the rotation curve of the mass model for
each stellar component scales as the square root of its mass-to-light ratio. 
There is no corresponding mass-to-light ratio for the gas of the interstellar medium as there is a well-understood relation 
between the observed flux at 21 cm and the mass of hydrogen atoms that emit it \citep{ISMbook}.
Consequently, the line for the gas components in Fig.\ \ref{fig:ugc2953ddo064} is practically fixed.

In addition to the mass-to-light ratio, there are two ``nuisance'' parameters that are sometimes considered in MOND fits:
distance and inclination. These are known from independent observations, but of course these have some uncertainty.
Consequently, the best MOND fit sometimes occurs for slightly different values of the distance and inclination, within
their observational uncertainties \citep{BBS,S1996,dBM98}. 

Distance matters because it sets the absolute scale.
The further a galaxy, the greater its mass for the same observed flux.
The distances to individual galaxies are notoriously difficult to measure.
Though usually not important, small changes to the distance can occasionally have powerful effects, especially in gas rich galaxies. 
Compare, for example, the fit to DDO 154 by \citet{LiRAR} to that of \citet{RKKYPRX}.

Inclinations matter because we must correct the observed velocities for the inclination of each galaxy as projected on the sky. 
The inclination correction is $V = V_{\mathrm{obs}}/\sin(i)$, so is small at large inclinations (edge-on) but large at small inclinations (face-on). 
For this reason, dynamical analyses often impose an inclination limit. This is an issue in any theory, but MOND is particularly
sensitive since $M \propto V^4$ so any errors in the inclination are amplified to the fourth power \citep[see Fig.\ 2 of][]{dBM98}.
Worse, inclination estimates can suffer systematic errors \citep{dBM98,verhTF,M12}: a galaxy seen face-on may have an oval 
distortion that makes it look more inclined than it is, but it can't be more face-on than face-on. 
  
MOND fits will fail if either the distance or inclination is wrong. 
Such problems cannot be discerned in fits with dark matter halos, which have ample flexibility to absorb the imparted 
variance \citep[see Fig.\ 6 of][]{dBM98}. Consequently, a fit with a dark matter halo will not fail if the distance happens to be wrong;
we just won't notice it. 

MOND generally fits rotation curves well \citep{K87,Milgrom1988,BBS,S1996,SV1998,dBM98,SMmond,gentileN1560,Gentile2011,swatersmond,LivRev,SSHM2013,AngusPoisson,AngusDiskMassMOND,Haghi2016,Hees2016,LiRAR,Sanders2019}.
There are of course exceptions \citep[e.g, NGC 2915][]{LiRAR}.
This is to be expected, as there are always some misleading data, especially in astronomy where it is impossible to control
for systematic effects in the same manner that is possible in closed laboratories. It is easily forgotten
that this type of analysis assumes circular orbits in a static potential, a condition that many spiral galaxies appear to have achieved
to a reasonable approximation but which certainly will not hold in all cases.

The best-fit mass-to-light ratios found in MOND rotation curve fits can be checked against independent stellar population models.
There is no guarantee that this procedure will return plausible values for the stellar mass-to-light ratio. 
Nevertheless, MOND fits recover the amplitude that is expected for stellar populations, the
expected variation with color, and the band-dependent scatter \citep[e.g., Fig.\ 28 of][]{LivRev}.
Indeed, to a good approximation, the rotation curve can be predicted directly from near-infrared data \citep{SV1998,M2020}
modulo only the inevitable scatter in the mass-to-light ratio.
This is a spectacular success of the paradigm that is not shared by dark matter fits \citep{K87,dBM97,dBBM03}.

Gas rich galaxies provide an even stronger test. 
When gas dominates the mass budget, the mass-to-light ratio of the stars ceases to have much leverage on
the fit. There is no 
fitting parameter for gas equivalent to the mass-to-light ratio for stars: the gas mass follows directly from the observations.
This enables MOND to predict the locations of such galaxies in the Baryonic Tully-Fisher plane \citep{M11}
and essentially their full rotation curves \citep{Sanders2019} with no free parameters \citep{M2020}.

\begin{table}[t]
\caption{Predictions of MOND}
 \label{tab:MONDpred}
\begin{center}
\begin{tabular}{l c}
\textbf{Prediction} & \textbf{Observation}  \\
Tully-Fisher Relation &   \\
~~$M_b$ alone specifies $V_f$ & + \\
~~Slope = 4 & + \\
~~No size or surface brightness residuals & + \\
$M_{\mathrm{dyn}}/L$ depends on surface brightness &  + \\
Central density relation  & +  \\
Rotation curve fits &   + \\
Stellar population mass-to-light ratios  & +
\end{tabular}
\end{center}
\end{table}

It should be noted that the acceleration scale \azero\ is kept fixed when fitting rotation curves.
If one allows \azero\ to vary, both it and the mass-to-light ratio spread over an unphysically large range of values \citep{LiRAR}.
The two are highly degenerate, causing such fits to be meaningless \citep{LiGN}: the data do not have the power to constrain
multiple parameters per galaxy. 

Table \ref{tab:MONDpred} lists the successful predictions of MOND that are discussed here.
A more comprehensive list is given by \citet{LivRev} and \citet{M2020} who also discuss some of the problems posed for dark matter. 
MOND has had many predictive successes beyond rotation curves \citep[e.g.,][]{MM13a,MM13b,Crater2} and has inspired
successful predictions in cosmology \citep[e.g.,][]{S1998M,M99,M00,S2001,CJP2015,M18}.
In this context, it makes sense to associate LSB galaxies with low density fluctuations in the initial conditions, 
thereby recovering the success of DD while its ills are cured by the modified force law. 
Galaxy formation in general is likely to proceed hierarchically but much more rapidly
than in \LCDM\ \citep{S2001,SK2001}, providing a natural explanation for both the age of stars in elliptical galaxies and allowing
for a subsequent settling time for the disks of spiral galaxies \citep{MONDgalform}.


The expert cosmologist may object that there is a great deal more data that must be satisfied. 
These have been reviewed elsewhere \citep{SMmond,bekenstein,LivRev,CJP2015} and are beyond the scope of this discussion. 
Here I note only that my experience has been that reports of MOND's falsification are greatly exaggerated.
Indeed, it has a great deal more explanatory power for a wider variety of phenomena than is generally appreciated \citep{MdB98a,MdB98b}. 

The most serious, though certainly
not the only, outstanding challenge to MOND is the dynamics of clusters of galaxies \citep{SMmond,angusbuote}. Contrary to the
case in most individual galaxies and some groups of galaxies \citep{Mondgroups,Mondgroups2}, MOND typically falls short of correcting 
the mass discrepancy in rich clusters by a factor of $\sim 2$ in mass. This can be taken as completely fatal, or as a being remarkably
close by the standards of astrophysics. Which option one chooses seems to be mostly a matter of confirmation bias: those who are quick
to dismiss MOND are happy to spot their own models a factor of two in mass, and even to assert that it is 
natural to do so \citep[e.g.,][]{Ludlow2017}. MOND is hardly alone in suffering problems with clusters of galaxies,
which also present problems for \LCDM\ \citep[e.g.,][]{angmcg,TooMuchLensing,ElGordoTooFat}.

A common fallacy seems to be that any failing of MOND is automatically considered to be support for \LCDM. 
This is seldom the case. 
More often than not, observations that are problematic for MOND are also problematic for \LCDM.
We do not perceive them as such because we are already convinced that non-baryonic dark matter must exist.
From that perspective, any problem encountered by \LCDM\ is a mere puzzle that will inevitably be solved,
while any problem encountered by MOND is a terminal failure of an irredeemably blasphemous hypothesis.
This speaks volumes about human nature but says nothing about how the universe works. 

\section{Concluding Remarks}
\label{sec:whylordwhy}

MOND has made many successful predictions. 
By any of the standards discussed by the philosophers of science, 
MOND has provided considerably more novel predictive content than any competing idea \citep[][]{Merrittbook}. 
This must teach us something. 

The most natural interpretation of the observations discussed here is that MOND is essentially correct, and is pointing the
way towards a deeper theory of dynamics \citep{milgromtheseproc}.
The obvious research program would seek to build on 
this \citep{TeVeS,skordis,milgromMI,bimetricMOND,bipolarDM,superfluidDM,Merritt_convention,FKP2018,SZ_sppedofgravity,SZ2020}.

From the perspective of our understanding of cosmology, the existence of non-baryonic dark matter is a seemingly unavoidable requirement,
so another approach is to attempt to explain dynamical data in that context. This does not occur naturally (\S \ref{sec:galform}),
and leads to a number of apparent self-contradictions \citep{M2020}.
Trying to explain this unexpected reality in terms of complicated feedback mechanisms seems like pounding the square peg
into the round hole: we're sure the answer has to be dark matter, so the misshapen peg must be made to fit. 
The anomaly for \LCDM\ is that MOND gets any prediction right, let alone so many.

The situation now is analogous to that in the time of Copernicus. 
We are piling complication upon complication to explain what at root is a simple phenomenon.
His words from nearly five centuries ago can be paraphrased for our current predicament: 
\begin{quotation}
``Those who devised the [\textit{eccentrics}/\textbf{feedback prescriptions}] seem thereby in large measure to have solved the problem of the apparent motions with appropriate calculations. But meanwhile they introduced a good many ideas which apparently contradict the first principles of [\textit{uniform motion}/\textbf{parsimony}]. Nor could they elicit or deduce from [\textit{the eccentrics}/\textbf{feedback effects}] the principal consideration, that is, the [\textit{structure of the universe and the true symmetry of its parts}/\textbf{ability to predict the kinematics of galaxies from their observed mass distribution}]. On the contrary, their experience was just like someone taking from various places hands, feet, a head, and other pieces, very well depicted, it may be, but not for the representation of a single person; since these fragments would not belong to one another at all, a monster rather than a man would be put together from them." 
--- Nicolaus Copernicus, \textit{De Revolutionibus} \citep[as translated by][]{Copernicus}, along with my paraphrase (bold face)
paralleling his original words (italicized). 
\end{quotation}

In the persistent absence of laboratory evidence to the contrary,
it remains possible that `dark matter' is a proxy for some deeper phenomenon, and our present conception of it is
nothing more than a hypothetical entity convenient to cosmic calculations. Like aether in the 19th century,  
cold dark matter is a substance that simply must exist given our present understanding of physics. But does it?

\section*{Acknowledgements}
This paper is based on an invited talk presented at the International Conference on Dark Matter \& Modified Gravity
hosted at RWTH Aachen University in February 2019, organized by the project ``LHC and Gravity'' within the interdisciplinary, 
DFG-funded research unit ``Epistemology of the LHC.'' I am grateful to Niels Martens for his work in leading the organization
of this event, and for his encouragement and support to participate. I am also grateful to many colleagues for stimulating discussions
about galaxy formation, dark matter, and MOND over the years, including Greg Bothun, Jim Schombert, Chris Mihos, Jim Peebles
Houjun Mo, Simon White, Donald Lynden-Bell, Thijs van der Hulst, Erwin de Blok, Bob Sanders, Vera Rubin, 
Jerry Sellwood, David Merritt, Moti Milgrom, Frank van den Bosch, Benoit Famaey, Pavel Kroupa, Dan Akerib, Tom Shutt, Glenn Starkman, 
Rachel Kuzio de Naray, Ji Hoon Kim, Federico Lelli, Marcel Pawlowski, and Pengfei Li, to name only a few. 
I thank Aviva Rothman for her perspective on Rosen's translation of \textit{De Revolutionibus}.
I am also grateful to Wayne Myrvold, the other editors, and two anonymous referees for very perceptive comments.
The images in Fig.\ \ref{fig:NGC3379N628NGC821} are based on photographic data obtained using the Oschin Schmidt Telescope 
on Palomar Mountain as part of the Palomar Observatory Sky Survey-II.
Digital versions of the scanned photographic plates were obtained for reproduction from the Digitized Sky Survey (archive.stsci.edu/dss). 
The Digitized Sky Surveys were produced at the Space Telescope Science Institute under U.S.\ Government grant NAG W-2166. 
The work of the author is supported in part by NASA ADAP grant 80NSSC19k0570 and NSF PHY-1911909.


\end{document}